\documentclass[aps,prl,final,twocolumn,letterpaper]{revtex4}

\usepackage{graphicx}   
\usepackage{import}                         
\usepackage{epstopdf}
\usepackage{amsmath} 
\usepackage{float} 
\usepackage{bm}
\usepackage{amssymb}
\usepackage{quotes}
\usepackage{indentfirst}
\usepackage{color}
\usepackage{transparent}
\usepackage{dcolumn}
\usepackage{braket}
\usepackage{multirow}
\usepackage{cancel} 
\usepackage{mdframed}
\usepackage{color}
\usepackage{bm}
\usepackage{dsfont}
\usepackage{slashed}
\usepackage{soul, color} 
\soulregister\ref{7}  
\soulregister\cite{7} 
\renewcommand{\st}[1]{}

\usepackage{xr}
\usepackage{mathtools}
\usepackage{natbib}

\begin{document}
\rmfamily


\title{Decoherence-free many-body Hamiltonians in nonlinear waveguide quantum electrodynamics}

\author{Aviv~Karnieli$^{1}$}
\email{karnieli@stanford.edu}
\author{Offek Tziperman$^{2}$}
\author{Charles~Roques-Carmes$^{1}$}
\author{Shanhui~Fan$^{1}$}

\affiliation{$^{1}$ E. L. Ginzton Laboratories, Stanford University, 348 Via Pueblo, Stanford, CA USA}
\affiliation{$^{2}$ Technion – Israel Institute of Technology, Haifa 32000, Israel}

\begin{abstract}
Enhancing interactions in many-body quantum systems, while protecting them from environmental decoherence, is at the heart of many quantum technologies. Waveguide quantum electrodynamics is a promising platform for achieving this, as it hosts infinite-range interactions and decoherence-free subspaces of quantum emitters. However, as coherent interactions between emitters are typically washed out in the wavelength-spacing regime hosting decoherence-free states, coherent control over the latter becomes limited, and many-body Hamiltonians in this important regime remain out of reach. Here we show that by incorporating emitter arrays with nonlinear waveguides hosting parametric gain, we obtain a unique class of many-body interaction Hamiltonians with coupling strengths that increase with emitter spacing, and persist even for wavelength-spaced arrays. We then propose to use these Hamiltonians to coherently generate decoherence-free states directly from the ground state, using only global squeezing drives, without the need for local addressing of individual emitters. Interestingly, we find that the dynamics approaches a unitary evolution in the limit of weak intra-waveguide squeezing, and discuss potential experimental realizations of this effect. Our results pave the way towards coherent control protocols in waveguide quantum electrodynamics, with applications including quantum computing, simulation, memory and nonclassical light generation.      
\end{abstract}


\maketitle
\textit{Introduction.} Many-body quantum entanglement is a crucial resource for a plethora of quantum technologies, including quantum sensing \cite{Degen2017QuantumSensing}, quantum simulation \cite{Georgescu2014QuantumSimulation} and quantum computation~\cite{MichaelA.NielsenandIsaacL.Chuang2010QuantumInformation}. Such schemes typically require strong interactions among system components, and the mitigation of environment-induced decoherence. One possible route for enhancing collective interactions, is by engineering the environment the system is coupled to \cite{Poyatos1996QuantumIons, Kienzler2015QuantumEngineering,Harrington2022EngineeredScience}. For example, driving an emitter system with a squeezed bosonic reservoir can exponentially enhance interaction strength and cooperativity \cite{Qin2018ExponentiallyAmplification,Leroux2018EnhancingCoupling,Zeytinoglu2017EngineeringVacuum,Burd2021QuantumInteractions}. To mitigate interactions with the environment and subsequent decoherence, it is possible to prepare such systems in a decoherence-free (DF) subspace \cite{Lidar1998Decoherence-FreeComputation, Zanardi1997NoiselessCodes}. The DF subspace hosts many-body entangled states that are decoupled from the environment, owing to destructive interference of their decay pathways.

A promising platform for realizing many-body quantum entanglement is waveguide quantum electrodynamics (WQED) \cite{Shen2005CoherentWaveguides,Shen2009TheoryAtom,Caneva2015QuantumFormalism,Asenjo-Garcia2017ExponentialArrays,Sheremet2023WaveguideCorrelations}, wherein one-dimensional arrays of quantum emitters are coupled to a waveguide, acting as a continuous photonic reservoir. The one-dimensional geometry gives rise to infinite-range interactions mediated by the waveguide mode~\cite{Sheremet2023WaveguideCorrelations}, which offers novel capabilities for quantum technologies \cite{Paulisch2016UniversalSubspaces, Malz2020NondestructiveQED, Tabares2023VariationalQED} and fundamental investigations of correlated phenomena \cite{Goban2015SuperradianceWaveguide,Zhang2019TheoryChain,Poshakinskiy2021QuantumInteractions, Tziperman2023TheEmission}. Reservoir engineering techniques in WQED utilizing a separately-prepared squeezed vacuum drive have been proposed \cite{You2018WaveguideVacuum,Bai2021GeneratingEngineering,Gutierrez-Jauregui2023DissipativeVacuum} for stabilizing steady-states using strong dissipation. Furthermore, as WQED can be used to realize the Dicke model of superradiance for wavelength-spaced arrays \cite{Sheremet2023WaveguideCorrelations}, it supports a DF subspace, as explored in several theoretical \cite{Albrecht2019SubradiantWaveguide, Paulisch2016UniversalSubspaces,Kockum2018Decoherence-FreeElectrodynamics, Pichler2015QuantumNetworks,Holzinger2022ControlQED} and experimental \cite{Zanner2022CoherentElectrodynamics} investigations.

Despite these advancements, enhancement and protection of many-body correlations in WQED is still challenging. Since the DF subspace is inherently decoupled from the waveguide mode, accessing it often requires the local addressing of individual emitters\cite{Paulisch2016UniversalSubspaces, Pichler2015QuantumNetworks, Holzinger2022ControlQED, Zanner2022CoherentElectrodynamics}, which can hamper scalability and fidelity \cite{Paulisch2016UniversalSubspaces}. Moreover, the coherent dynamics of the emitters is intrinsically limited in the wavelength-spacing regime hosting DF states, in which the many-body Hamiltonian vanishes \cite{Albrecht2019SubradiantWaveguide}. This holds even if reservoir engineering schemes are employed \cite{You2018WaveguideVacuum,Bai2021GeneratingEngineering,Gutierrez-Jauregui2023DissipativeVacuum}, as the latter often rely on dissipation rather than on unitary evolution. Ultimately, these limitations hinder the realization of unitary dynamics and many-body Hamiltonians in this important regime of WQED.


In this Letter, we lift these limitations by considering quantum emitter arrays coupled to \textit{nonlinear} $\chi^{(2)}$ waveguides, serving as traveling-wave parametric amplifiers (or squeezers) \cite{Macklin2015AAmplifier,Qiu2023BroadbandAmplifier,Nehra2022Few-cycleNanophotonics}. Unlike squeezed-reservoir engineering, the parametric gain accumulation along the nonlinear waveguide in our system gives rise to an unconventional coherent atomic interaction that \textit{increases} with inter-atomic distance, and persists even in the wavelength-spacing regime. Surprisingly, we find that in the limit of weak intra-waveguide squeezing, the dynamics approaches an adiabatic unitary evolution under a many-body Hamiltonian, and the system can be coherently and globally driven to DF excited states directly from its ground state, without the need for local addressing of the emitters. We also investigate the emergent polaritonic excitations in the DF subspace, and outline possible experimental WQED platforms. Our findings pave the way towards coherent control protocols in WQED, with applications ranging from quantum computation \cite{Paulisch2016UniversalSubspaces}, memory \cite{Asenjo-Garcia2017ExponentialArrays} and simulation \cite{Tabares2023VariationalQED} to nonclassical light sources \cite{Gonzalez-Tudela2015DeterministicDissipation,Tziperman2023TheEmission}.


\begin{figure}
    \centering
    \includegraphics[scale = 0.26]{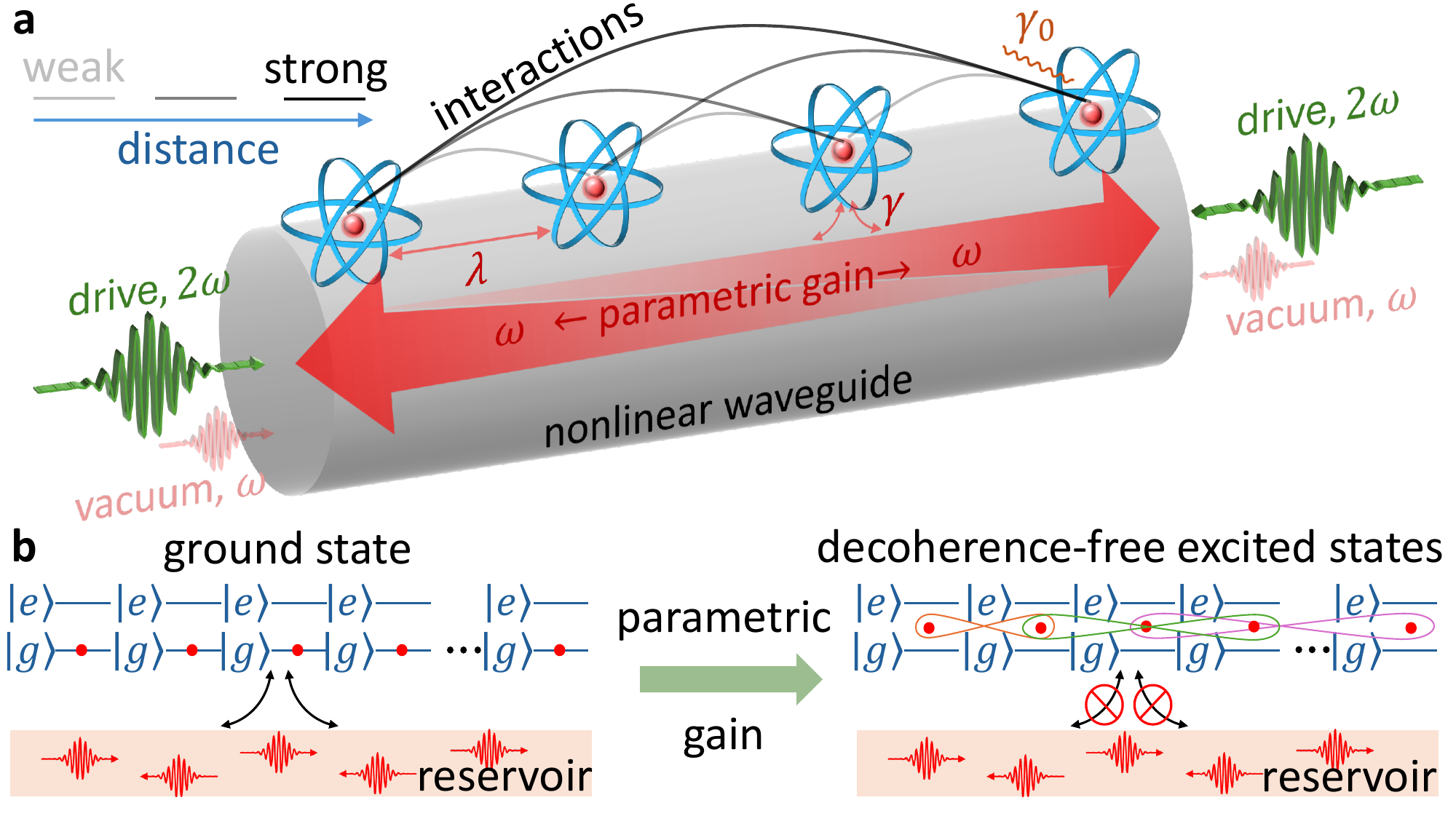}
    \caption{\textbf{Nonlinear waveguide quantum electrodynamics (WQED).} \textbf{a} An array of $N$ emitters is coupled to a $\chi^{(2)}$ nonlinear waveguide with coupling strength $\gamma$ and decay into free space $\gamma_{\mathrm{0}}$. The waveguide is driven by a classical field at $2\omega$, generating broadband parametric gain around frequency $\omega$. Even when the emitters are evenly spaced by a wavelength, the parametric gain can still mediate coherent interactions between them, having the unusual property of increased interaction strength with increasing inter-atomic distance. \textbf{b} The parametric gain can coherently and globally drive the system from its ground state to decoherence-free excited states.   }
   \label{fig:concept}
\end{figure}

\textit{WQED with parametric gain.} Our system comprises an array of $N$ identical atoms with transition frequency $\omega$, equally-spaced by a distance $\Delta z$ and coupled to a nonlinear waveguide of length $L$, as depicted in Fig. (1a). The position of atom $j$ is $z_j = z_1 + (j-1)\Delta z$ (where $0\leq z_1\leq z_j\leq z_{N}\leq L$), having a lowering (raising) operator $\sigma_{-,j}$ ($\sigma_{+,j}$). The waveguide is simultaneously driven by two counter-propagating classical pumps at a frequency of $2\omega$, driving a degenerate squeezing interaction that generates right- and left-propagating squeezed vacuum fields that gradually build up (amplified) along the propagation direction, with frequency $\omega$, wavevector $k$, and group velocity $v_g$. We assume that the atoms are all coupled to the waveguide with equal rate $\gamma$. The atoms can also have an external decay rate as a result of coupling to reservoirs other than the waveguide ($\gamma_0$ in Fig. 1a), leading to a coupling efficiency $\beta\equiv\gamma/(\gamma+\gamma_\mathrm{0}) < 1$ \cite{Sheremet2023WaveguideCorrelations}. We shall first consider the ideal case $\beta =1$, and later discuss realistic experimental platforms with $\beta < 1$.

Denoting the interaction-picture field envelope operators by $A_{s}(z,t)$, with $s=\rightleftarrows$, we show in the Supplementary Material (SM) sections (S1-S2), that the fields at atomic positions $z_j$ and time $t$ can be written in terms of a Bogoliubov transformation of the retarded input fields $A_{\rightarrow}(0,t-z_j/v_g)$ and $A_{\leftarrow}(L,t-(L-z_j)/v_g)$, assumed to be in the vacuum state. Under the Markovian approximation (assuming the atomic chain is not too long such that retardation could be neglected $v_g/L\gg\gamma$ \cite{Caneva2015QuantumFormalism}), the dynamics of the atomic system can be obtained using the SLH formalism \cite{Combes2017TheNetworks}. The system evolves under the Lindblad master equation $\dot{\rho} = \mathcal{L}[\rho]$, with the Lindbladian superoperator defined as $\mathcal{L}[\rho] = -i/\hbar[H,\rho]+\sum_{s=\rightleftarrows} \mathcal{D}_{L_s}[\rho] $, with $\mathcal{D}_{X}[\rho] =  X \rho X^{\dagger} -1/2\lbrace X^{\dagger}X,\rho\rbrace$ denotes the dissipator corresponding to a jump operator $X$.  

The most general form of the resulting Hamiltonian and jump operators is given in the SM section (S3). Hereafter, we make the following assumptions: (i) wavelength spacing of the array $\Delta z = m\lambda$ with $m$ being an integer and $\lambda=2\pi/k$ \endnote{similar conclusions hold under the more general Bragg condition $\Delta z = m\lambda/2$}. (ii) $z_1 =0$ and $z_N = L$, i.e., the atom array and the nonlinear region of the waveguide coincide. (iii) symmetric squeezing drives: same squeezing gain and phase for the right- and left-propagating fields. Denoting by $r$ and $\theta$ the total squeeze parameter and phase, respectively, the accumulated squeeze parameters from the right and left, up to atom $j$, are written respectively as $r_{\rightarrow,j} = r(j-1)/(N-1)$ and $r_{\leftarrow,j} = r(N-j)/(N-1)$. Under these assumptions, we find

\begin{equation}
\label{SLH Hamiltonian and jumps}
    \begin{split}
    &H = \frac{\gamma }{2}\sum_{i,j=1}^{N} \sinh{\left( \frac{r|i-j|}{N-1}\right)}  \left( e^{-i\theta} \sigma_{+,j}\sigma_{+,i} + e^{i\theta}\sigma_{-,j}\sigma_{-,i}\right), \\ & L_s = \sqrt{\gamma}\sum_{j=1}^N\left(\cosh{r_{s,j}} \sigma_{-,j}-ie^{-i\theta} \sinh{r_{s,j}} \sigma_{+,j}\right), 
    \end{split}
\end{equation}

\begin{figure}
    \centering
    \includegraphics[trim={0.14cm 10.6cm 0 0},clip, scale = 0.43]{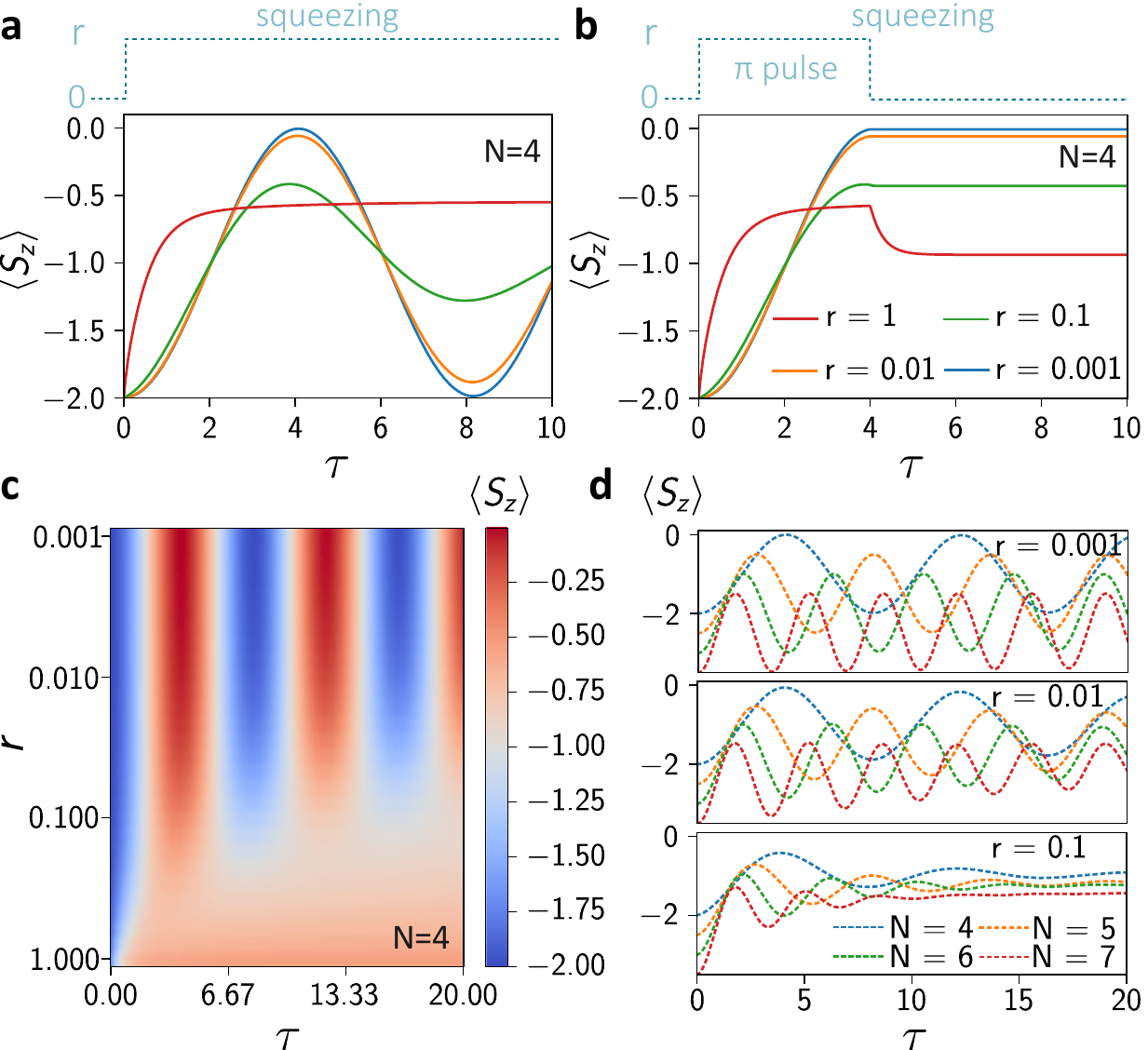}
    \caption{\textbf{Emergence of coherent dynamics under weak parametric gain.} The evolution of the global population is plotted for different squeezing parameters $r$ and emitter number $N$. Whereas for $r=1$ the dynamics is completely dissipative, coherent Rabi oscillations emerge for $r\ll 1$. \textbf{a-b} $N=4$ atoms. For $r\ll 1$, the population remains excited even after the squeezing is turned off (e.g., when the squeezing drive is a $\pi$-pulse as in \textbf{b}), and does not decay back to the reservoir: a signature of decoherence-free states. \textbf{c} Oscillations for $N=4$ atoms and a continuous change in squeezing parameter. \textbf{d} Oscillations for different atom numbers $N$ and varying squeezing levels, where the Rabi frequency increases with $N$. All plots assume $\beta=1$. }
   \label{fig:concept}
\end{figure}

Interestingly, the interaction Hamiltonian in Eq. (\ref{SLH Hamiltonian and jumps}) persists even though the atoms are spaced by a wavelength. In contrast, in the absence of parametric gain in the waveguide under the same conditions, coherent inter-atomic interactions are washed out and the dynamics is purely-dissipative (the Dicke-superradiant regime) \cite{Albrecht2019SubradiantWaveguide}. The Hamiltonian demonstrates an unusual property of interaction strength that is \textit{exponentially increasing} with increased inter-atomic distance. This peculiar feature stems from the amplification of photons mediating the interaction between distant atoms. While exponential-range coupling was previously considered in atomic simulators with tree-like geometry \cite{Periwal2021ProgrammableClouds, Bentsen2019TreelikeAtoms}, our Hamiltonian involves an all-to-all connectivity, courtesy of the waveguide mode. Furthermore, unlike the conventional spin-exchange interaction, here each pairwise interaction term has the form of a two-axis twisting Hamiltonian with a separation-dependent coupling \cite{Ma2011QuantumSqueezing,Liu2011SpinTwisting}. Thus, the proposed system can realize even more diverse forms of spin squeezing and quantum simulation.

To assess the dynamical features of our system in the wavelength-spacing regime, in Figs. 2a-c we plot the expectation value of the global population operator $S_z = \sum_i \sigma_{z,i}$ for $N=4$ quantum emitters starting from the global ground state ($\braket{S_z}=-N/2$), for different levels of squeezing. When $r\sim 1$, the system is quickly stabilized to a steady state through dissipation. However, when $r\ll 1$, coherent Rabi oscillations emerge, displaying increasing visibility for weaker squeezing. When varying the squeezing parameter, we rescaled the time to $\tau = \gamma rt$. As we show below, $\gamma r$ is the universal scale of the emergent adiabatic oscillations, which become slower for decreasing $r$. The continuous transition between complete dissipation and coherent oscillations as a function of $r$ and $\tau$ is plotted in Fig. 2c, and Fig. 2d displays Rabi oscillations for $4\leq N \leq 7$ and select values of $r$, where Rabi frequency increases with $N$. Richer dynamics for $N\geq 8$, comprising several Rabi frequencies, is presented in SM, section S7. Interestingly, as shown in Fig. 2b, if the squeezing drive is turned off, the emitter population does not decay: this is a manifestation of the excitation of DF states.

\begin{figure}
    \centering
    \includegraphics[trim={0.2cm 1.5cm 0 0},clip, scale = 0.43]{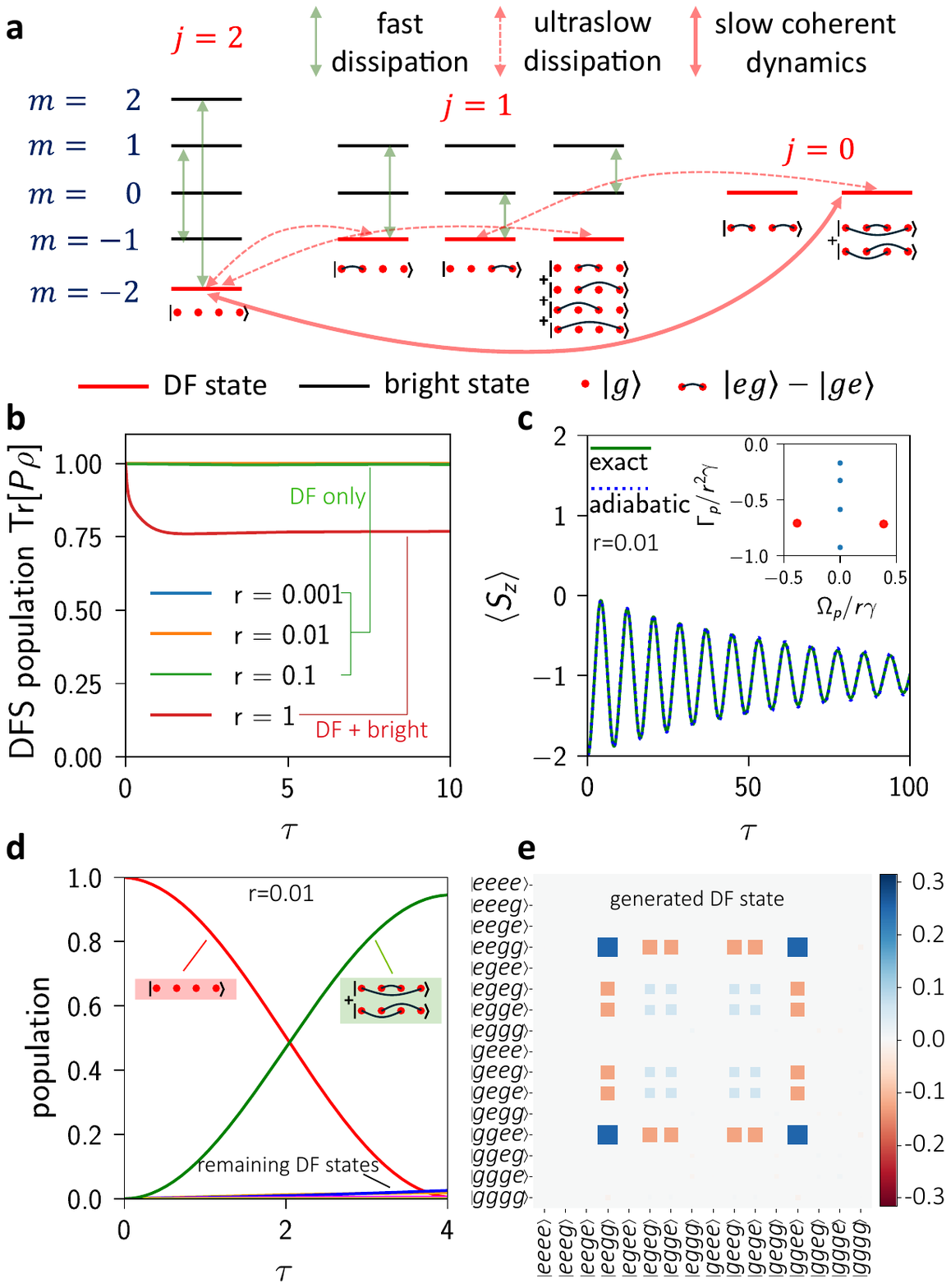}
    \caption{\textbf{Coherent adiabatic dynamics in the decoherence-free subspace.} \textbf{a} Total angular momentum subspaces $j$ (here illustrated for 4 atoms). Each $j$-subspace hosts DF states with total $z$-component $m=-j$, comprising antisymmetric dimer states \cite{Poshakinskiy2021DimerizationElectrodynamics}. The parametric gain breaks permutation symmetry, thus coupling different $j$ subspaces. Fast dissipation of bright states $\sim \gamma$ can then be adiabatically eliminated, leaving slow coherent dynamics between DF states with rate $\sim r\gamma$, as well as ultraslow dissipation $\sim r^2\gamma$. \textbf{b} Projection of the state $\rho$ onto the DF subspace $\mathcal{H}_{\mathrm{P}}$ indeed shows that for $r\ll 1$, all populated states are decoherence-free. \textbf{c} Exact and adiabatically-eliminated dynamics, resembling a finite-lifetime polariton. Inset: eigenvalues of $H_{\mathrm{eff}}$ on the complex plain (polariton energies highlighted). \textbf{d} DF state populations along a half Rabi cycle, converting between the global ground state and the entangled $j=0$ DF state with a fidelity of 94.5\% (which improves with decreasing $r$). \textbf{e} Density matrix of the generated state in \textbf{d}. All plots assume $\beta = 1$. }
   \label{fig:concept}
\end{figure}

\textit{Decoherence-free subspace and coherent adiabatic dynamics.} In the Dicke regime of conventional WQED, permutation symmetry allows to express the dynamics in terms of the collective operators $S_{\pm}=\sum_i \sigma_{\pm,i}$ and $S_z$. The commutation of $S^2 = S_z^2 + \lbrace S_+, S_- \rbrace/2$ with the Hamiltonian and all jump operators then ensures \cite{Albert2014SymmetriesEquations} that the Lindbladian can be block-diagonalized into subspaces of total angular-momentum $j=0,(1/2), 1, (3/2),..., N/2$ for even (odd) $N$, as depicted in Fig. 3a. The DF subspace contains all states that are annihilated by $S_- $, which are the (possibly degenerate) states of the form $m=-j$, where $-j\leq m\leq j$ is the $z$-component quantum number of the total angular momentum. 

In the presence of permutation symmetry, DF states do not couple with one another, as they reside in different $j$-subspaces. In particular, the global ground state \endnote{experimentally, the system can be prepared in the global ground state by thermalization, laser cooling or coherent initialization} $\ket{gg...g}$ with $j=N/2$ and $m=-N/2$ is decoupled from other excited states ($j=-m>N/2$) in the DF subspace. The nonlinear parametric gain allows one to break permutation symmetry [as in Eq. (\ref{SLH Hamiltonian and jumps})], without the need for local addressing of individual emitters \cite{Paulisch2016UniversalSubspaces, Pichler2015QuantumNetworks, Holzinger2022ControlQED, Zanner2022CoherentElectrodynamics}. As such, different $j$-subspaces can now be dynamically coupled via global driving, and as a result, so do DF states.   

In fact, in the weak-squeezing limit $r\ll1$ of Eq. (\ref{SLH Hamiltonian and jumps}), it can be shown that the dynamics involves \textit{only} DF states. As can be seen in Fig. 3b, for $r\ll 1$ the state resides exclusively in the DF subspace. To understand this, consider the projection operator onto the DF subspace, $\mathcal{H}_P$, denoted as $P$ (and its super-operator version $\mathcal{P}$, such that $\mathcal{P}\rho = P\rho P$). Writing an arbitrary state $\rho$ as $\rho = \mathcal{P}\rho + (1-\mathcal{P})\rho$, and inspecting the dynamics of $\mathcal{P}\rho$ and $(1-\mathcal{P})\rho$, we show in the SM section (S5) that bright states $(1-\mathcal{P})\rho$ experience superradiant decay with rate $\sim \gamma N$ back into the DF subspace. On the other hand, we find that the dynamics of DF states $\mathcal{P}\rho$ occurs at a much slower rate $\sim r\gamma \ll N\gamma$. 

Remarkably, we further find that the evolution of the DF states can be arbitrarily close to a unitary for $N\geq 4$ atoms. As we show in the SM section (S5), the dissipative part of the slow dynamics within $\mathcal{H}_P$ occurs at ultraslow rates $\sim r^2\gamma$, whereas the coherent part evolves with a rate $\sim r\gamma$ (see arrows in Fig. 3a). This means that we should be expecting Rabi oscillations with a relatively-long coherence time, with $\sim 1/2r$ visible periods. Thus, in the weak squeezing limit, and to first order in $r$, the dynamics approaches a unitary evolution (as can be seen in Fig. 2), wherein $H \to PHP$, corresponding to quantum Zeno dynamics \cite{Paulisch2016UniversalSubspaces,Harrington2022EngineeredScience}. For the case of two and three atoms, $PHP=0$, such that the dynamics is completely dissipative (with rate $\sim r^2\gamma$).

These observations motivate the adiabatic elimination \cite{Finkelstein-Shapiro2020AdiabaticSystems} of the fast dynamics of bright states, leaving us with an adiabatic Hamiltonian and jump operators $H_{\mathrm{ad}},~\lbrace L_{\mathrm{ad},i}\rbrace$ that act on states in $\mathcal{H}_P$ alone. We outline the adiabatic elimination procedure in the SM section (S6), which for general open quantum systems can be performed numerically in terms of Liouville superoperators. We find that our adiabatically-eliminated dynamics hosts polaritonic excitations within the DF subspace, which are the eigenstates of the non-Hermitian effective Hamiltonian $H_{\mathrm{eff}} = H_{\mathrm{ad}} - i \frac{1}{2}\sum_i L_{\mathrm{ad},i}^{\dagger}L_{\mathrm{ad},i}$. These polaritonic eigenstates may in general have complex eigenfrequencies $\Omega_p +i \Gamma_p$, with $\Gamma_p \leq 0$, where $\Omega_p \sim r\gamma$ and $\Gamma_p \sim r^2 \gamma$. Fig. 3c shows the time evolution of $\langle S_z\rangle$ under the exact [Eq. (\ref{SLH Hamiltonian and jumps})] and adiabatically-eliminated dynamics for $N=4$, where the inset shows the complex eigenvalues of the DF-subspace polaritons. While polaritons in the DF subspace have a finite coherence time, owing to ultraslow dissipation and dephasing in $\mathcal{H}_p$, we stress that when the squeezing drive is turned off (at any finite time $t_0$), so does this dissipation; one is left with a DF state that does not decay for $t\geq t_0$ (see Fig. 2b). In this manner, we show in Figs. 3d-e that one can adiabatically generate DF excited states directly from the global ground state. Such states can be highly-entangled as can be seen in the Hinton diagram of the corresponding density matrix in Fig. 3e.

\begin{figure}
    \centering
    \includegraphics[trim={0.1cm 20cm 0 0},clip, scale = 0.42]{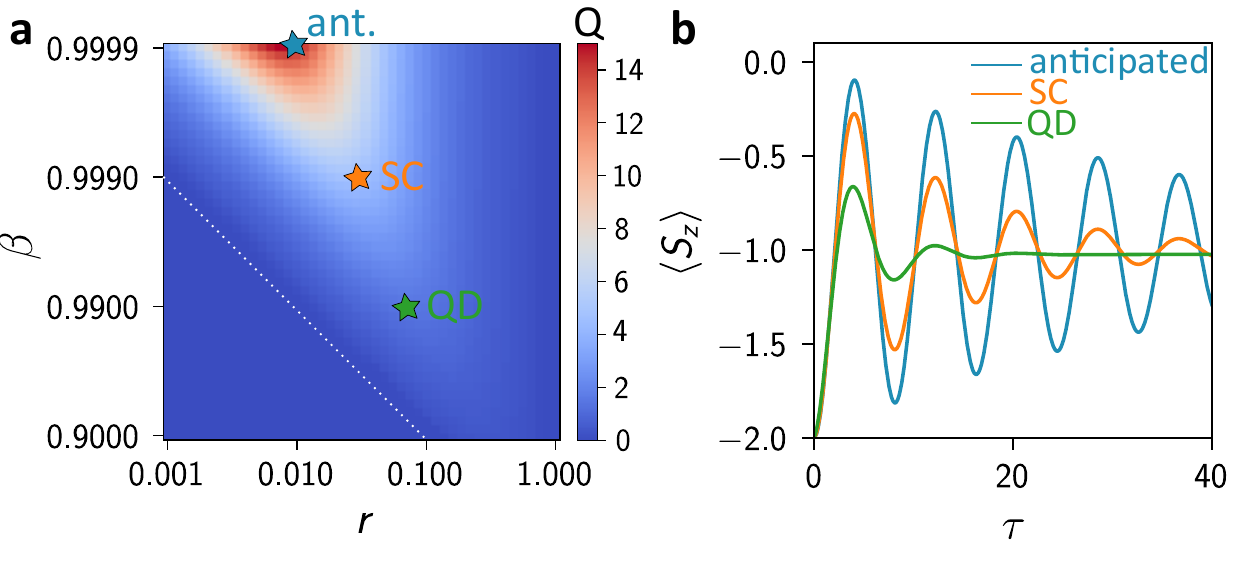}
    \caption{\textbf{Experimental prospects in realistic WQED platforms.} Coherent adiabatic oscillations $\sim r\gamma$ must be faster than the radiative decay rate $\gamma_{\mathrm{rad}} = [(1-\beta)/\beta] \gamma$, ideally with $1-\beta \ll r $. \textbf{a} Polaritonic quality factor $Q=\Omega_{\mathrm{p}}/2\Gamma_{\mathrm{p}}$ for $N=4$ atoms. Onset of oscillations $1-\beta = r$ shown in dashed white line. State of the art values of $\beta = 0.99$ for quantum dots (QD) \cite{Foster2019TunableWaveguide,Sheremet2023WaveguideCorrelations} and $\beta = 0.999$ superconducting qubits (SC) \cite{Mirhosseini2019CavityMirrors,Sheremet2023WaveguideCorrelations} are shown at the maximal $Q$ value. \textbf{b} Dynamics with finite $1-\beta$ values corresponding to the marked points in \textbf{a}: $\beta=0.99,0.999,0.9999$ and $r=0.086, 0.025, 0.01$, respectively.}
   \label{fig:concept}
\end{figure}

\textit{Experimental considerations.} While DF states are immune to dissipation into the waveguide reservoir, they are still susceptible to radiative loss outside the waveguide (and other forms of decoherence and dephasing). Further, as with other adiabatic schemes in WQED \cite{Paulisch2016UniversalSubspaces}, one needs to ensure that the total decoherence rate is slower than the rate of the adiabatic dynamics. In our case, observing the predicted Rabi oscillations of DF states amounts to requiring that $r\ll 1$ and $1-\beta \ll r$, which implies near-ideal coupling efficiency $\beta \simeq 1$. 

In Fig. 4a we plot the DF polariton quality factors $Q=|\Omega_p/2\Gamma_p|$ as a function of squeezing parameter $r$ and coupling efficiency $\beta$. The onset of Rabi oscillations (the line $1-\beta = r$) is also shown. Considering the state-of-the-art values for two different WQED platforms \cite{Sheremet2023WaveguideCorrelations}, quantum dots (QD), with $\beta =0.99$ \cite{Arcari2014Near-UnityWaveguide,Foster2019TunableWaveguide,Sheremet2023WaveguideCorrelations} and superconducting (SC) qubits with $\beta =0.999$ \cite{Mirhosseini2019CavityMirrors,Sheremet2023WaveguideCorrelations}, we plot in Fig. 4b the corresponding dynamics with the ideal $r$ value for each case as indicated in Fig. 4a. Anticipating future improvements of coupling efficiency, we also plot the nearly-ideal case of $\beta = 0.9999$ and $r=0.01$ ($1-\beta \sim r^2$), for which loss is commensurate with the ultraslow DF polariton dissipation. Realization of our scheme in the SC and QD platforms can be done by integrating the emitters into travelling-wave parametric amplifiers. With the advent of SC Josephson amplifiers \cite{Macklin2015AAmplifier, Grimsmo2017SqueezingAmplifiers} and integrated nonlinear optical waveguides \cite{Nehra2022Few-cycleNanophotonics}, we believe that experimental proof-of-concept is within reach.

\textit{Discussion and outlook.} We have shown that coupling atom arrays to nonlinear waveguides hosting parametric gain, leads to a unique coherent inter-atomic interaction in a regime where such interactions are typically washed out. We have further shown how this dynamics approaches unitary evolution within the DF subspace, and can be used to generate entangled DF states directly from the ground state using only global drives. 

Unlike dissipative reservoir engineering schemes with squeezed vacuum \cite{You2018WaveguideVacuum,Bai2021GeneratingEngineering,Gutierrez-Jauregui2023DissipativeVacuum} (see also SM section S4), our proposal utilizes an emergent coherent many-body Hamiltonian interaction to dynamically steer desired atomic states. As the coherent dynamics becomes richer for larger atom numbers (see SM section S7), further work is needed to discover universal coherent control schemes over arbitrarily-large DF-states using global parametric driving. One possible route to achieve this is to employ time-dependent \cite{Nishad2023QuantumArrays} and spectrally-shaped \cite{Hurvitz2023Frequency-domainInformation} drives, as well as globally-controlled waveguide dispersion \cite{Yanik2004StoppingTransparency}. The gradual formation of such DF states can be then probed by photon scattering \cite{Shi2013Two-photonEffect, Shi2018UltrastrongTheory, Ke2019InelasticStates, Asenjo-Garcia2017ExponentialArrays, Holzinger2022ControlQED}. It would be intriguing to also consider other forms of nonlinearity leading to exotic physics in WQED, as proposed using Kerr cavity arrays \cite{Wang2020SupercorrelatedWaveguides, Wang2024Long-RangeWaveguides}. 

The unique nonlocal structure of our interaction Hamiltonian Eq. (\ref{SLH Hamiltonian and jumps}) may prove useful for implementing quantum simulations \cite{Periwal2021ProgrammableClouds, Bentsen2019TreelikeAtoms}, topological squeezing-based models \cite{Wan2023Quantum-Squeezing-InducedEffect}, as well as spin-squeezing with distance-dependent coupling \cite{Perlin2020SpinInteractions}. The possibility to drive DF states can be useful in quantum thermodynamic applications in WQED \cite{Lu2022Steady-StateDriving}, such as energy storage in quantum batteries \cite{Quach2020UsingBatteries}. Finally, selective releasing of excitations stored in the DF states back into the waveguide may enable novel schemes for on-demand, macroscopic nonclassical light sources \cite{Asenjo-Garcia2017ExponentialArrays,Tziperman2023TheEmission}.

\section{Funding} A.K. is supported by the VATAT-Quantum fellowship by the Israel Council for Higher Education; the Urbanek-Chodorow postdoctoral fellowship by the Department of Applied Physics at Stanford University; the Zuckerman STEM leadership postdoctoral program; and the Viterbi fellowship by the Technion. C.~R.-C. is supported by a Stanford Science Fellowship. S. F. acknowledges the support from the Department of Energy. (Grant No. DE-FG02-07ER46426).  
\section{Acknowledgments} The authors acknowledge fruitful discussions with Janet Zhong, Renwen Yu and Nicholas Rivera.
\section{Disclosures} The authors declare no conflict of interest.
\section{Data availability} Data may be obtained from the authors upon reasonable request. 
\bibliography{references}
\bibliographystyle{ieeetr}

\end{document}


\rmfamily

\title{Supplementary Material: Decoherence-free many-body Hamiltonians in nonlinear waveguide quantum electrodynamics}

\author{Aviv~Karnieli$^{1}$}
\email{karnieli@stanford.edu}
\author{Offek Tziperman$^{2}$}
\author{Charles~Roques-Carmes$^{1}$}
\author{Shanhui~Fan$^{1}$}

\affiliation{$^{1}$ E. L. Ginzton Laboratories, Stanford University, 348 Via Pueblo, Stanford, CA USA}
\affiliation{$^{2}$ Technion – Israel Institute of Technology, Haifa 32000, Israel}

\maketitle

\tableofcontents

\newpage

\section{Nonlinear waveguide quantum electrodynamics}
We begin by writing our system Hamiltonian as $H=H_0 + H_1$ where
\begin{equation}
\label{H0}
    \begin{split}
        H_0 = \hbar\sum_{s=\rightleftarrows}\Bigg[ \int_{-\infty}^{\infty} dz A_s^{\dagger}(z)\left(-id_s v_g \frac{\partial}{\partial z}\right)A_s (z)
        + \kappa_{s} \int_{0}^L A_s(z) A_s(z) dz + \kappa_{s}^* \int_0^L  A_s^{\dagger}(z) A_s^{\dagger}(z) dz  \Bigg],
    \end{split}
\end{equation}
acts only on the waveguide field and describes the parametric gain present in the region $0\leq z\leq L$, where $A_s$, $s=\rightleftarrows$ denotes the field envelope annihilation operators, $v_g$ denotes the magnitude of the group velocity, $\kappa_s$ is the nonlinear coupling (proportional to the pump field and material nonlinearity $\chi^{(2)}$) arising from the squeezing interaction and $d_{\rightarrow}=1, d_{\leftarrow} = -1$. The field operators obey the canonical commutation relation for bosons:
\begin{equation}
    \label{eq:commutation}
    \left[ A_s(z), A_s^\dagger(z') \right] = \delta(z-z').
\end{equation}
In writing Eq. (\ref{H0}), we assumed that the squeezing interaction is phase-matched ($k_{2\omega}-2k_{\omega}=0$). The second term in the total Hamiltonian $H$,  
\begin{equation}
\label{H1}
    \begin{split}
        H_1 = \hbar\sqrt{2\pi }g \sum_{s=\rightleftarrows}\sum_{j=1}^{N} \big[e^{d_s i kz_j}A_s (z_j)\sigma_{+,j} + e^{-d_s i kz_j}A_s^{\dagger} (z_j)\sigma_{-,j}\big],
    \end{split}
\end{equation}
corresponds to the interaction of the field with the atoms in the rotating wave approximation, where $g$ is the coupling constant of an atom to the field (assumed real and uniform for all atoms), $z_j $ and $\sigma_{-(+),j}$ are the position and lowering (raising) operator, respectively, of atom $j$, with $0\leq z_1\leq z_j\leq z_{N}\leq L$, and $k$ is the carrier wavevector of the guided field. In the manuscript we assume equal spacing of the emitters $z_j=z_1 + (j-1) \Delta z$, while in principle different spacing can be used to engineer the phase and magnitude of the effective coupling between emitters.

\section{Interaction picture}

We can write the Heisenberg equations of motion for the fields evolving under Hamiltonian $H_0$ of Eq. (\ref{H0}):
\begin{equation}
\label{Heisenberg}
    \begin{split}
        i\hbar \frac{\partial}{\partial t} A_s(z) = -i d_s \hbar v_g \frac{\partial}{\partial z} A_s(z) + 2\hbar \kappa_s^* \mathrm{rect} \left(\frac{z}{L}\right) A_s^{\dagger}(z)  
    \end{split}
\end{equation}
where $\mathrm{rect} \left(x\right) =1$ if $0\le x\le 1$ and 0 otherwise. From now on we will omit the $z$ functional dependence of the operators.
Denoting $r_s = 2|\kappa_s|L/v_g$ and $\theta_s=\mathrm{arg}[\kappa_s]$ we can write
\begin{equation}
\label{Heisenberg_vectorized}
    \begin{split}
        \left(\frac{\partial}{\partial t}+d_s v_g\frac{\partial}{\partial z}\right)\begin{pmatrix}A_s \\ A_s^{\dagger}\end{pmatrix} = \frac{r_s v_g}{L} \mathrm{rect}\left(\frac{z}{L}\right)\begin{pmatrix} 0 & -ie^{-i\theta_s} \\ ie^{i\theta_s} & 0 \end{pmatrix} \begin{pmatrix}A_s \\ A_s^{\dagger}\end{pmatrix}
    \end{split}
\end{equation}
Now, we can transform our fields according to

\begin{equation}
\label{Heisenberg_B_transformation}
    \begin{split}
        \begin{pmatrix}B_{s,+} \\ B_{s,-}\end{pmatrix} = \frac{1}{\sqrt{2}}\begin{pmatrix}
            ie^{i\theta_s} & 1 \\ -ie^{i\theta_s} & 1 
        \end{pmatrix} \begin{pmatrix}A_s \\ A_s^{\dagger}\end{pmatrix} 
    \end{split}
\end{equation}
which allows us to diagonalize the dynamics as
\begin{equation}
\label{Heisenberg_diagonalized}
    \begin{split}
        \left(\frac{\partial}{\partial t}+d_s v_g\frac{\partial}{\partial z}\right)\begin{pmatrix}B_{s,+} (z,t)\\ B_{s,-}(z,t)\end{pmatrix} = \frac{v_g}{L} \mathrm{rect}\left(\frac{z}{L}\right)\begin{pmatrix} r_s & 0 \\ 0 & -r_s \end{pmatrix} \begin{pmatrix}B_{s,+} (z,t)\\ B_{s,-}(z,t)\end{pmatrix}
    \end{split}
\end{equation}
Next, moving to the frequency domain (with $f(t)=\int d\omega e^{-i\omega t} F(\omega) $), we find:
\begin{equation}
\label{Heisenberg_diagonalized_frequency}
    \begin{split}
        \frac{\partial}{\partial z}\begin{pmatrix}B_{s,+}(z,\omega) \\ B_{s,-}(z,\omega)\end{pmatrix} = d_s \begin{pmatrix} i\frac{\omega}{v_g} + \mathrm{rect}\left(\frac{z}{L}\right) \frac{r_s}{L}  & 0 \\ 0 & i\frac{\omega}{v_g} - \mathrm{rect}\left(\frac{z}{L}\right) \frac{r_s}{L} \end{pmatrix} \begin{pmatrix}B_{s,+}(z,\omega) \\ B_{s,-}(z,\omega)\end{pmatrix}
    \end{split}
\end{equation}
The solutions are
\begin{equation}
\label{Heisenberg_B_solution_frequency}
    \begin{split}
        \begin{pmatrix}B_{s,+}(z,\omega) \\ B_{s,-}(z,\omega)\end{pmatrix} = e^{id_s \frac{\omega}{v_g}(z-z_{0,s})} \begin{pmatrix} e^{r_{s}\xi_{s}(z)}  & 0 \\ 0 & e^{-r_{s}\xi_{s}(z)} \end{pmatrix} \begin{pmatrix}B_{s,+}(z_{0,s},\omega) \\ B_{s,-}(z_{0,s},\omega)\end{pmatrix}
    \end{split}
\end{equation}
where $z_{0,s}$ are $z_{0,\rightarrow}=0$ and $z_{0,\leftarrow}=L$ are the boundaries for the right- and left-propagating fields, respectively, and
\begin{equation}
\label{Heisenberg_xi}
    \begin{split}
        \xi_{\rightarrow}(z) = \int_{0}^{z/L}\mathrm{rect}(x) dx = \begin{cases}
            0,  &z\leq 0\\
            z/L, & 0\leq z\leq L\\
            1, & z\geq L
        \end{cases}, ~~~~\xi_{\leftarrow}(z)=\xi_{\rightarrow}(L-z).
    \end{split}
\end{equation}
Note that $\xi_s(z_{0,s})=0$ and that the solutions satisfy the boundary conditions at $z=z_{0,s}$. Moving back to the time domain we have
\begin{equation}
\label{Heisenberg_B_solution_time}
    \begin{split}
        \begin{pmatrix}B_{s,+}(z,t) \\ B_{s,-}(z,t)\end{pmatrix} = \begin{pmatrix} e^{r_{s}\xi_{s}(z)}  & 0 \\ 0 & e^{-r_{s}\xi_{s}(z)} \end{pmatrix} \begin{pmatrix}B_{s,+}(z_{0,s},t-d_s \frac{z-z_{0,s}}{v_g}) \\ B_{s,-}(z_{0,s},t-d_s \frac{z-z_{0,s}}{v_g})\end{pmatrix}
    \end{split}
\end{equation}
and, in terms of the original fields,
\begin{equation}
\label{Heisenberg_A_solution_time}
    \begin{split}
        \begin{pmatrix}A_{s}(z,t) \\ A_{s}^{\dagger}(z,t)\end{pmatrix} = \begin{pmatrix} \cosh[r_{s}\xi_{s}(z)]  & -ie^{-i\theta_s}\sinh[r_{s}\xi_{s}(z)] \\ i e^{i\theta_s}\sinh[r_{s}\xi_{s}(z)] & \cosh[r_{s}\xi_{s}(z)] \end{pmatrix} \begin{pmatrix}A_{s}(z_{0,s},t-d_s \frac{z-z_{0,s}}{v_g}) \\ A_{s}^{\dagger}(z_{0,s},t-d_s \frac{z-z_{0,s}}{v_g})\end{pmatrix}
    \end{split}
\end{equation}
Now let us consider the fields at the atomic positions $0\leq z_i \leq L$ and define the input noise operators as 
\begin{equation}
\label{Gamma_operators}
    \Gamma_{i}^s(t) =\sqrt{v_g} A_s\left(z_{0,s},t-d_s\frac{z_i-z_{0,s}}{v_g}\right)
\end{equation}
then, using $\xi_{\rightarrow}(z_i) = z_i/L,~~\xi_{\leftarrow}(z_i) = 1-z_i/L$, we can write 
\begin{equation}
\begin{split}
\label{Fields_at_atom_position}
    &\sqrt{v_g}A_{\rightarrow}(z_i,t) = \cosh(r_{\rightarrow} z_i/L)\Gamma_{i}^{\rightarrow}(t) -ie^{-i\theta_{\rightarrow}}\sinh(r_{\rightarrow}z_i/L)\Gamma_{i}^{\rightarrow\dagger}(t) \\
    &\sqrt{v_g}A_{\leftarrow}(z_i,t) = \cosh[r_{\leftarrow}(1- z_i/L)]\Gamma_{i}^{\leftarrow}(t) -ie^{-i\theta_{\leftarrow}}\sinh[r_{\leftarrow}(1-z_i/L)]\Gamma_{i}^{\leftarrow\dagger}(t)
\end{split}
\end{equation}
  
Finally, we consider the atom-field interaction, Eq. (\ref{H1}), and transform $H_1$ to the interaction picture by substituting the expressions for $A_s(z,t)$ in terms of $\Gamma_j^s(t)$, arriving at
\begin{equation}
\label{HI}
    \begin{split}
    H_I(t) = e^{iH_0 t} H_1 e^{-iH_0 t} = \hbar \sqrt{\gamma} \sum_{s=\rightleftarrows}\sum_{j=1}^{N} c_j^{s\dagger} \Gamma_j^s(t) + c_j^s \Gamma_j^{s\dagger}(t)
    \end{split}
\end{equation}
where $\gamma = 2\pi g^2/v_g$ and where we defined a new set of system operators, which inherit the Bogoliubov transformation:
\begin{equation}
\label{c_j}
    \begin{split}
    c_j^s &= e^{-id_s \phi j}\cosh{r_{s,j}}\sigma_{-,j} -i e^{-i\theta_s} e^{id_s \phi j}\sinh{r_{s,j}}\sigma_{+,j}. 
    \end{split}
\end{equation}
In Eq. (\ref{c_j}), $\phi = k\Delta z$ is the inter-atom phase accumulation, and $r_{s,j}$ with $s=\rightleftarrows$ stands for the accumulated parametric gain (squeezing) up to atom $j$, for each of the right- and left-propagating fields. Explicitly, we can write $r_{\rightarrow,j} = \bar{r}_{\rightarrow} + (j-1)\Delta r_{\rightarrow}$ and $r_{\leftarrow,j} = \bar{r}_{\leftarrow} + (N-j)\Delta r_{\leftarrow}$ for $j=1,2,...,N$. Both $\bar{r}_{s}$ and $\Delta r_{s}$ could be expressed in terms of the gain (squeezing) per unit length $\mathcal{G}_s=2|\kappa_s|/v_g$ as $\bar{r}_{\rightarrow} =\mathcal{G}_{\rightarrow}z_1,~~\bar{r}_{\leftarrow} =\mathcal{G}_{\leftarrow}(L-z_{N})$ and $\Delta r_s = \mathcal{G}_s\Delta z$. In fact, \ $\bar{r}_{s}$ corresponds to the squeeze parameters of the right and left going modes of \textit{squeezed vacuum reservoirs}, prepared in the segments $0\leq z\leq z_1$ and $z_{N}\leq z\leq L$, i.e., prior to their interaction with the emitter array; whereas the $\Delta r_s$ correspond to \textit{squeezing accumulation} in between neighboring atoms. Thus, our model can capture two relevant limits: the squeezed reservoir-engineering limit, previously considered in the literature \cite{You2018WaveguideVacuum,Bai2021GeneratingEngineering,Gutierrez-Jauregui2023DissipativeVacuum}, is formally obtained by taking $\mathcal{G}_s \to 0$ and $z_1, L-z_{N} \to \infty$, such that $\Delta r_s\to 0$ while $\bar{r}_{s}$ can remain finite. The opposite limit, which we consider in this work, is of pure squeezing accumulation, where $z_1 =0$ and $z_{N} =L$, with $\bar{r}_{s}=0$ and finite $\Delta r_s$. 

We note that to obtain the compact form of Eq.~(16), we factored-out physically-meaningless global phases from $c_j^{\rightarrow}$ and $c_j^{\leftarrow}$, and absorbed constant phase factors into the squeezing phases $\theta_{\rightarrow}$ and $\theta_{\leftarrow}$, which we assume to be freely tunable experimentally.

\section{SLH formalism}
The SLH formalism  \cite{Combes2017TheNetworks} has been developed for theoretical calculations of input-output networks, systems in which the output of a quantum system is fed as the input to another quantum system. We would like to transform our Hamiltonian into a form that is compatible with the SLH formalism. To achieve this, we move to the frequency domain by defining frequency-dependent operators $b_{\omega}^s=A_s(z_{0,s},\omega)$, such that the noise operators can be written as $\Gamma_j^s(\omega)=e^{i\omega\tau_j^s}b_{\omega}^s$, with $\tau_j^{\rightarrow} = z_j/v_g$ and $\tau_j^{\leftarrow} = (L-z_j)/v_g$, we find
\begin{equation}
\label{HI_frequency}
    \begin{split}
    H_I =  \sum_{s=\rightleftarrows}\int d\omega \left(\hbar \omega b_{\omega}^{s\dagger}b_{\omega}^s + \hbar \sqrt{\gamma}\sum_{j=1}^{N} c_j^{s\dagger} e^{i\omega\tau_j^s}b_{\omega}^s + c_j^s e^{-i\omega\tau_j^s}b_{\omega}^{s\dagger}\right)
    \end{split}
\end{equation}
 This form of the Hamiltonian allows us to use the SLH cascading rules \cite{Combes2017TheNetworks} in the limit $\tau_{j}^s\ll \gamma^{-1}$. In terms of the SLH notation, we identify the SLH triple for system $i$ coupled to bath $s=\rightleftarrows$ as
\begin{equation}
\label{SLH_i}
    \begin{split}
    S_i^s=1,~~L_{i}^s = \sqrt{\gamma}c_{i}^s,~~H_i^s =0
    \end{split}
\end{equation}
or $G_{i}^s = (1,L_i^s,0)$. According to the SLH rules, the left- and right-going systems are first cascaded:
\begin{equation}
\begin{split}
    \label{SLH_cascading}
    &G^{\rightarrow} = G^{\rightarrow}_N\triangleleft G^{\rightarrow}_{N-1}\triangleleft ...\triangleleft G^{\rightarrow}_{1}\\
    &G^{\leftarrow} = G^{\leftarrow}_1\triangleleft ...\triangleleft G^{\leftarrow}_{N-1}\triangleleft G^{\leftarrow}_{N}
    \end{split}
\end{equation}
Then, the total system is concatenated:
\begin{align}
\label{SLH_concatenation}
    G = G^{\rightarrow}\boxplus G^{\leftarrow}
\end{align}
The Hamiltonian and jump operators resulting from this procedure are then
\begin{equation}
\label{SLH_Hamiltonian}
    H=\sum_{i=1}^N \sum_{j=1}^{N} \Theta(j-i)\frac{L_j^{\rightarrow\dagger}L_i^{\rightarrow}-L_i^{\rightarrow\dagger}L_j^{\rightarrow}}{2i} + \Theta(i-j)\frac{L_j^{\leftarrow\dagger}L_i^{\leftarrow}-L_i^{\leftarrow\dagger}L_j^{\leftarrow}}{2i}
\end{equation}
and
\begin{equation}
\label{SLH_jumps}
    L^{\rightarrow}=\sum_{i=1}^N L_i^{\rightarrow},~~L^{\leftarrow}=\sum_{i=1}^N L_i^{\leftarrow}
\end{equation}
where the system evolves according to the master equation

\begin{equation}
\label{master_equation}
    \dot{\rho} = -i[H,\rho] + \sum_{s=\rightleftarrows} \mathcal{D}_{L^s} [\rho]
\end{equation}
with
\begin{equation}
\label{dissipator}
    \mathcal{D}_{L^s}[\rho] = L^s\rho L^{s\dagger}-\frac{1}{2} L^{s\dagger}L^s\rho-\frac{1}{2} \rho L^{s\dagger}L^s
\end{equation}
denoting the dissipator of jump operator $L^s$. We note that the same master equation is obtained by taking the Born-Markov approximation procedure \cite{Manzano2020AEquation,Scully1997}.

We now wish to explicitly write the Hamiltonian and jump operators in terms of the system parameters. For simplicity we assume that the right- and left-going squeezing parameters are equal such that $\bar{r}_{\rightarrow}=\bar{r}_{\leftarrow}\equiv\bar{r}$ and $\Delta r_{\rightarrow}=\Delta r_{\leftarrow}\equiv\Delta r$, but allow the squeezing phases $\theta_{\rightleftarrows}$ to be different, as well as a nonzero $\bar{r}$ (squeezed reservoir engineering). We find:

\begin{equation}
\label{SLH_Hamiltonian_explicit}
\begin{split}
    H&=\frac{\gamma}{2}\sum_{i=1}^N\sum_{j=1}^N \sin(\phi |j-i|)\cosh(\Delta r|j-i|)[\sigma_{+j}\sigma_{-,i}+\sigma_{+i}\sigma_{-,j}] \\& +\frac{\gamma}{2}\sum_{i=1}^N\sum_{j=1}^N \sinh(\Delta r|j-i|)\lbrace\Theta(j-i)[e^{-i\theta_{\rightarrow}}e^{i\phi (i+j)}\sigma_{+j}\sigma_{+,i}+e^{i\theta_{\rightarrow}}e^{-i\phi (i+j)}\sigma_{-i}\sigma_{-,j}] \\&+\Theta(i-j)[e^{-i\theta_{\leftarrow}}e^{-i\phi (i+j)}\sigma_{+j}\sigma_{+,i}+e^{i\theta_{\leftarrow}}e^{i\phi (i+j)}\sigma_{-i}\sigma_{-,j}]\rbrace  
\end{split}
\end{equation}
and
\begin{align}
\label{SLH_jumps_explicit_right}
L^{\rightarrow} &= \sqrt{\gamma} \sum_j e^{-i\phi j} \cosh[\bar{r}+\Delta r (j-1)]\sigma_{-,j}-ie^{i\phi j}e^{-i\theta_{\rightarrow}}\sinh[\bar{r}+\Delta r (j-1)]\sigma_{+,j}\\
\label{SLH_jumps_explicit_left}
L^{\leftarrow} &= \sqrt{\gamma} \sum_j e^{i\phi j} \cosh[\bar{r}+\Delta r (N-j)]\sigma_{-,j}-ie^{-i\phi j}e^{-i\theta_{\leftarrow}}\sinh[\bar{r}+\Delta r (N-j)]\sigma_{+,j}
\end{align}

To further simplify we consider the case of wavelength spacing $\phi =2\pi$ and equal squeezing phases $\theta_{\rightarrow}=\theta_{\leftarrow}\equiv\theta$, to find:

\begin{equation}
\label{SLH_Hamiltonian_explicit_simplified}
\begin{split}
    H&=\frac{\gamma}{2}\sum_{i=1}^N\sum_{j=1}^N \sinh(\Delta r|j-i|)[e^{-i\theta}\sigma_{+j}\sigma_{+,i}+e^{i\theta}\sigma_{-i}\sigma_{-,j}]  
\end{split}
\end{equation}
and
\begin{align}
\label{SLH_jumps_explicit_right_simplified}
L^{\rightarrow} &= \sqrt{\gamma} \sum_j \cosh[\bar{r}+\Delta r (j-1)]\sigma_{-,j}-ie^{-i\theta}\sinh[\bar{r}+\Delta r (j-1)]\sigma_{+,j}
\\
L^{\leftarrow} &= \sqrt{\gamma} \sum_j \cosh[\bar{r}+\Delta r (N-j)]\sigma_{-,j}-ie^{-i\theta}\sinh[\bar{r}+\Delta r (N-j)]\sigma_{+,j} \label{SLH_jumps_explicit_left_simplified}
\end{align}
Eq. (1) of the main text is recovered by setting $\bar{r}=0$ (equivalently, setting $z_1=0$ and $z_N=L$), such that no squeezed reservoir is input to the system. This assumption significantly simplifies possible experimental implementations of our proposal. 

\section{Decoherence free subspace}

Decoherence-free (DF) states are states that satisfy

\begin{equation}
    S_-\ket{\psi_{DF}}=0
\end{equation}
i.e. they form the null space of the global lowering operator $S_-=\sum_j \sigma_{-,j}$. What is unique about DF states is that if the system obeys superradiant dynamics
\begin{equation}
    \dot{\rho} = \gamma \mathcal{D}_{S_-}[\rho] = S_-\rho S_+ +\frac{1}{2}\lbrace S_+S_-,\rho \rbrace
\end{equation}
then $\mathcal{D}_{S_-}[\ket{\psi_{DF}}\bra{\psi_{DF}}]=0$. That is, DF states do not decay under superradiant dynamics. Using block-diagonalization of the global $S_z$ and $S^2$ operators on the basis of angular-momentum-like states:
\begin{align}
    & S_z \ket{j,m,\alpha} = m\ket{j,m,\alpha} \\
    & S^2 \ket{j,m,\alpha} = j(j+1)\ket{j,m,\alpha}
\end{align}
where $\alpha$ is a degeneracy index \cite{Paulisch2016UniversalSubspaces}, one can write the DF states as
\begin{equation}
    \ket{\psi_{DF}}= \ket{j,-j,\alpha}
\end{equation}

\section{Comparison between squeezed reservoir engineering and accumulated squeezing}

We now use our model Eqs. (\ref{SLH_Hamiltonian_explicit_simplified})-(\ref{SLH_jumps_explicit_left_simplified}) to compare the familiar scheme of squeezed reservoir engineering (with $\bar{r}\neq 0,~\Delta r = 0$) \cite{Qin2018ExponentiallyAmplification,Leroux2018EnhancingCoupling,Zeytinoglu2017EngineeringVacuum,Burd2021QuantumInteractions} to our proposal of squeezing accumulation ($\bar{r}=0,~\Delta r\neq 0$) in the Bragg regime. Setting (without loss of generality) $\theta=-\pi/2$, we have for the reservoir engineering (RE) scheme that
\begin{equation}
\begin{split}
\label{Reservoir_engineering}
    &H_{\mathrm{RE}}=0\\
    &L^{\rightarrow}_{\mathrm{RE}} = L^{\leftarrow}_{\mathrm{RE}} = \sqrt{\gamma}\left(  \cosh\bar{r}S_-+\sinh \bar{r}S_{+}\right)
\end{split}
\end{equation}
where $S_{\pm}=\sum_j\sigma_{\pm,j}$, while for the squeezing accumulation (SA) scheme, we have that
\begin{equation}
\begin{split}
\label{Squeezing_accumulation}
    &H_{\mathrm{SA}}=i\frac{\gamma}{2}\sum_{i=1}^N\sum_{j=1}^N \sinh(\Delta r|j-i|)[\sigma_{+j}\sigma_{+,i}-\sigma_{-i}\sigma_{-,j}]  \\
& L^{\rightarrow}_{\mathrm{SA}} = \sqrt{\gamma} \sum_j \cosh[\Delta r (j-1)]\sigma_{-,j}+\sinh[\Delta r  (j-1)]\sigma_{+,j}\\ 
&L^{\leftarrow}_{\mathrm{SA}} = \sqrt{\gamma} \sum_j \cosh[\Delta r (N-j)]\sigma_{-,j}+\sinh[\Delta r (N-j)]\sigma_{+,j}
\end{split}
\end{equation}
Note that since $H_{\mathrm{RE}}=0$, the RE dynamics in the Dicke regime is completely dissipative. Further, since $\lbrace L^s_{\mathrm{RE}}\rbrace$ are permutationally symmetric (and both commute with $S^2$), the dynamics cannot couple different total angular momentum subspaces (the $j$-subspaces of Fig. 3 of the main text), and thus cannot couple between different decoherence free (DF) states. 

In contrast, for the SA scheme, the dynamics can still have a coherent part in the Dicke regime, since $H_{\mathrm{SA}}\neq 0$. Further, as both $H_{\mathrm{SA}}$ and $\lbrace L^s_{\mathrm{SA}}\rbrace$ break permutation symmetry (and do not commute with $S^2$), they can couple different $j$-subspaces, and by extension, between different DF states. In fact, as we show in the next section, in the weak squeezing limit $\Delta r \ll 1$ the dynamics between DF states can be arbitrarily close to a unitary evolution.

\section{Emergence of coherent dynamics between states in the decoherence-free subspace}
Denoting $\Delta r=r/(N-1)$, where $r$ is the total accumulated squeeze parameter, and without loss of generality choosing $\theta = -\pi/2$, we take the weak squeezing limit $r\ll 1$ of Eq. (1) of the main text [alternatively, of Eq. (\ref{Squeezing_accumulation}) above], to find 
\begin{equation}
\label{H_L_linearized}
\begin{split}
    &H=i r\frac{\gamma}{2}\sum_{i=1}^N\sum_{j=1}^N \frac{|j-i|}{N-1}(\sigma_{+j}\sigma_{+,i}-\sigma_{-i}\sigma_{-,j})  \\
& L^{\rightarrow} = \sqrt{\gamma} \left(S_-+ rJ_+^{\rightarrow}\right)\\ 
&L^{\leftarrow} = \sqrt{\gamma}\left( S_- + rJ_+^{\leftarrow} \right)
\end{split}
\end{equation}
where $J_{\pm}^{\rightarrow} = \sum\frac{j-1}{N-1}\sigma_{\pm,j}$ and $J_{\pm}^{\leftarrow} = \sum\frac{N-j}{N-1}\sigma_{\pm,j}$ are collective weighted operators that arise from the (linearized) inter-atomic parametric gain accumulation in the waveguide. Note that $J_{\rightarrow,\pm}+J_{\leftarrow,\pm} = S_{\pm}$.

We first show that the dynamics of DF states evolves over much slower timescales than that of bright states. To show this, we consider a state $\rho_{\mathrm{DF}}$ inside the DF subspace $\mathcal{H}_P$, and a bright state $\rho_{\mathrm{B}}$. We want to find the timescales over which $\rho_{\mathrm{DF}}$ and $\rho_{\mathrm{B}}$ evolve (and, perhaps, mix) according to the system Linbladian $\mathcal{L}[\rho] = -i[H,\rho] +\sum_{s=\rightleftarrows} \mathcal{D}_{L^s}[\rho]$.  First, we find for bright states that
\begin{equation}
\label{Bright_Lindbladian}
\mathcal{L}[\rho_{\mathrm{B}}] = 2\gamma\mathcal{D}_{S_-} [\rho_{\mathrm{B}}] +  O(r\gamma)
\end{equation}
where $\mathcal{D}_{S_-}$ is the superradiant dissipator. Considering the structure of the total angular momentum $j$-subspaces of Fig. 3 of the main text, every such bright state will eventually decay to the lowest angular momentum projection value $m=-j$, which is no other than a DF state. Thus, we conclude that bright states experience superradiant decay on a fast timescale $\sim (N\gamma)^{-1}$ back into the DF subspace.

In contrast, using the fact that the DF subspace is the nullspace of the global lowering operator such that $S_- \rho_{\mathrm{DF}} = \rho_{\mathrm{DF}} S_+ = 0$, we find
\begin{equation}
\label{DF_Lindbladian}
\begin{split}
    \mathcal{L}[\rho_{\mathrm{DF}}] &= -i[H,\rho_{\mathrm{DF}}]-\frac{\gamma}{2} r \sum_{s=\rightleftarrows}(S_+J_+^s \rho_{\mathrm{DF}} +\rho_{\mathrm{DF}} J_-^sS_-) + \gamma r^2 \sum_{s=\rightleftarrows}\mathcal{D}_{J^s_+}[\rho_{\mathrm{DF}}] \\& = -i[H,\rho_{\mathrm{DF}}]-\frac{\gamma }{2} r\sum_{s=\rightleftarrows}\lbrace S_+ J_+^s + J_-^s S_-,\rho_{\mathrm{DF}}\rbrace + \gamma r^2 \sum_{s=\rightleftarrows}\mathcal{D}_{J^s_+}[\rho_{\mathrm{DF}}] \\& = -i[H,\rho_{\mathrm{DF}}]-\frac{\gamma }{2} r\lbrace S_+ S_+ + S_-S_-,\rho_{\mathrm{DF}}\rbrace + \gamma r^2 \sum_{s=\rightleftarrows}\mathcal{D}_{J^s_+}[\rho_{\mathrm{DF}}] \\& = -i(\tilde{H}\rho_{\mathrm{DF}} -\rho_{\mathrm{DF}}\tilde{H}^{\dagger}) + \gamma r^2 \sum_{s=\rightleftarrows}\mathcal{D}_{J^s_+}[\rho_{\mathrm{DF}}]
\end{split}
\end{equation}
where we used $J_{\pm}^{\rightarrow}+J_{\pm}^{\leftarrow} = S_{\pm}$ and defined the non-Hermitian Hamiltonian
\begin{equation}
\label{Nonhermitian_Hamiltonian}
\begin{split}
\tilde{H} &= H - ir \frac{\gamma}{2}(S_+S_+ +S_-S_-) \\&= i r\frac{\gamma}{2}\left[ \sum_{i=1}^N\sum_{j=1}^N \frac{|j-i|}{N-1}(\sigma_{+j}\sigma_{+,i}-\sigma_{-i}\sigma_{-,j}) - (S_+S_+ +S_-S_-)\right]    
\end{split}
\end{equation}
meaning that the dynamics of the DF states occurs on a much slower timescale $\sim r\gamma \ll \gamma$.

These observations motivate the adiabatic elimination of the fast dynamics of bright states. We detail the adiabatic elimination procedure in the next section, and focus here on the main conclusions. First, we have shown that every "leakage" of state population from the DF subspace into the subspace of bright states will rapidly decay back to the DF subspace. If we start at any state inside the DF subspace, the adiabatic dynamics ensures that we stay in the DF subspace. This can be checked numerically by plotting the DF subspace population $\mathrm{Tr}[P\rho]$, which approaches unity as $r\ll 1$ (see Fig. 3b of the main text). 

However, such population leakage processes can still, in principle, cause decoherence and dephasing within the DF subspace. To assess this effect, we consider the projection operator onto the DF subspace, denoted by $P$, and its superoperator form $\mathcal{P}$, such that $\mathcal{P}\rho = P\rho P$. The complementary projection superoperator onto the bright subspace is denoted by $\mathcal{Q}=1-\mathcal{P}$, such that $\mathcal{Q}\rho = Q\rho P + P\rho Q + Q\rho Q$, with $Q=1-P$. Every state can be decomposed into $\rho =\mathcal{P}\rho + \mathcal{Q}\rho$, and every superoperator can be written as $\mathcal{O} = \mathcal{P}\mathcal{O}\mathcal{P} + \mathcal{P}\mathcal{O}\mathcal{Q}+\mathcal{Q}\mathcal{O}\mathcal{P} + \mathcal{Q}\mathcal{O}\mathcal{Q}$. Thus, we can decompose the Lindbladian according to how it maps each of the state components onto the two possible subspaces:
\begin{equation}
\label{Coupling_Lindbladians}
\begin{split}
    \mathcal{L}_{\mathrm{DF}\to\mathrm{DF}}[\rho] &= \mathcal{P}\mathcal{L}\mathcal{P}\rho 
    \\ \mathcal{L}_{\mathrm{DF}\to\mathrm{B}}[\rho] &= \mathcal{Q}\mathcal{L}\mathcal{P}\rho
    \\ \mathcal{L}_{\mathrm{B}\to\mathrm{DF}}[\rho] &= \mathcal{P}\mathcal{L}\mathcal{Q}\rho
    \\ \mathcal{L}_{\mathrm{B}\to\mathrm{B}}[\rho] &= \mathcal{Q}\mathcal{L}\mathcal{Q}\rho
\end{split}
\end{equation}
Luckily, we already know how the Lindbladian acts on $\rho_{\mathrm{DF}} = \mathcal{P}\rho$ and $\rho_{\mathrm{B}} = \mathcal{Q}\rho$, through Eqs. (\ref{Bright_Lindbladian})-(\ref{DF_Lindbladian}). Calculating the corresponding projections, we find
\begin{equation}
\label{Coupling_Lindbladians_explicit}
\begin{split}
    \mathcal{L}_{\mathrm{DF}\to\mathrm{DF}}[\rho_{\mathrm{DF}}] &= -i(P \tilde{H} P \rho_{\mathrm{DF}} - \rho_{\mathrm{DF}} P\tilde{H}^{\dagger} P) + O(r^2\gamma) 
    \\ \mathcal{L}_{\mathrm{DF}\to\mathrm{B}}[\rho_{\mathrm{DF}}] &= -i (Q \tilde{H} P \rho_{\mathrm{DF}} - \rho_{\mathrm{DF}} P\tilde{H}^{\dagger} Q) + O(r^2\gamma)
    \\ \mathcal{L}_{\mathrm{B}\to\mathrm{DF}}[\rho_{\mathrm{B}}] &= 2\gamma P\mathcal{D}_{S_-} [\rho_{\mathrm{B}}] P +  O(r\gamma)
    \\ \mathcal{L}_{\mathrm{B}\to\mathrm{B}}[\rho_{\mathrm{B}}] &= 2\gamma\left( Q\mathcal{D}_{S_-} [\rho_{\mathrm{B}}] P +  P\mathcal{D}_{S_-} [\rho_{\mathrm{B}}] Q +  Q\mathcal{D}_{S_-} [\rho_{\mathrm{B}}] Q\right)  +  O(r\gamma)
\end{split}
\end{equation}
where $\tilde{H}$ is the non-Hermitian Hamiltonian of Eq. (\ref{Nonhermitian_Hamiltonian}). Specifically, we have that 
\begin{equation}
\label{Coupling_Hamiltonians_explicit}
\begin{split}
    P\tilde{H}P &= PHP
    \\ Q\tilde{H}P &= QHP -ir \frac{\gamma}{2} Q(S_+S_+ +S_-S_-)P
\end{split}
\end{equation}
We can now envision our dynamics as a Hermitian system (the DF subspace) with Hamiltonian $H_{\mathrm{sys}} = PHP\sim r\gamma$, coupled to a discrete reservoir (bright states) through a Hamiltonian coupling $H_{\mathrm{sys,R}}=Q\tilde{H}P + P\tilde{H}^{\dagger}Q$, with characteristic rate $\Omega \sim r\gamma$, and where the reservoir can also evolve and decay back to the system through rapid and purely dissipative transitions with rate $\Gamma \sim \gamma$. Finally, we can employ the known result \cite{Reiter2012EffectiveSystems,Finkelstein-Shapiro2020AdiabaticSystems} in such settings in the limit $\Gamma \gg \Omega$  wherein the system's decoherence rate owing to the population transfer from the system to the reservoir with Hamiltonian coupling scales as $\Omega^2/\Gamma \sim r^2 \gamma$. The derivation can be done either using perturbation theory in the small parameter $\Omega/\Gamma$ \cite{Reiter2012EffectiveSystems} or by adopting a superoperator approach \cite{Finkelstein-Shapiro2020AdiabaticSystems}. The details can be found in these references and we will not repeat the derivation here.

Remarkably, this result means that in the limit $r\ll 1$, the adiabatic dynamics approaches a \textit{coherent} evolution under the Hamiltonian $PHP\sim r\gamma$, since all decoherence rates [within the DF subspace (dissipator in Eq. (\ref{DF_Lindbladian})) or due to population leakage outside it] scale as $\sim r^2 \gamma$. The new Hamiltonian $PHP$ is exactly the one emergent from quantum Zeno dynamics \cite{Paulisch2016UniversalSubspaces,Harrington2022EngineeredScience}.

\section{Adiabatic elimination procedure}
Using the same notations from the previous section, we follow the procedure outlined in Ref. \cite{Finkelstein-Shapiro2020AdiabaticSystems} to numerically evaluate the adiabatically-eliminated Lindbladian in Liouville space:
\begin{equation}
\label{adiabatic_elimination}
    \mathcal{L}_{\mathrm{ad}}=\mathcal{U}[\mathcal{P}\mathcal{L}\mathcal{P}-\mathcal{P}\mathcal{L}\mathcal{Q}(\mathcal{Q}\mathcal{L}\mathcal{Q})^{-1}\mathcal{Q}\mathcal{L}\mathcal{P}]\mathcal{U}^{\dagger}
\end{equation}
where we have also introduced a unitary change of basis $\mathcal{U}$ to the DF basis states, such that $\mathcal{U}\rho = U\rho U^{\dagger}$. To find the Lindbladian in the original space of density matrices, we compute numeriaclly the Kraus operators \cite{Manzano2020AEquation} $M_i$ that correspond to the differential CPTP map generated by $\mathcal{L}_{\mathrm{ad}}$:
\begin{equation}
\label{CPTP_map}
    \rho(t+dt) = \mathcal{E}[\rho (t)]=\sum_{i=1}^K M_i \rho (t) M_i^{\dagger}=e^{\mathcal{L}_{\mathrm{ad}}dt}\rho(t)
\end{equation}
The Kraus operators are well known to satisfy \cite{Manzano2020AEquation}
\begin{equation}
\label{Kraus}
\begin{split}
    &M_0 = I-\left(iH_{\mathrm{ad}} + \frac{1}{2}\sum_{i=1}^K L_{\mathrm{ad},i}^{\dagger}L_{\mathrm{ad},i}\right) dt\\
    &M_i =  L_{\mathrm{ad},i}\sqrt{dt}
\end{split}
\end{equation}
The adiabatically-eliminated Hamiltonian $H_{\mathrm{ad}}$ together with the jump operators $L_{\mathrm{ad},i}$ generate a new Lindblad master equation in the DF subspace, with Lindbladian:
\begin{equation}
\label{adiabatically_eliminated_Lindbladian}
    \mathcal{L}_{\mathrm{ad}} [\rho] = -i[H_{\mathrm{ad}},
    \rho] + \sum_{i=1}^K \mathcal{D}_{\mathrm{ad},i} [\rho]
\end{equation}

In particular we shall be interested in the effective, non-Hermitian Hamiltonian
\begin{equation}
\label{effective_Hamiltonian}
    H_{\mathrm{eff}} = H_{\mathrm{ad}} - i\frac{1}{2} \sum_{i=1}^K L_{\mathrm{ad},i}^{\dagger}L_{\mathrm{ad},i}
\end{equation}
which we find to host polariton-like excitations, formed by DF states. Such eigenstates of $H_{\mathrm{eff}}$ have complex eigenvalues $\Omega_{\mathrm{p}} + i\Gamma_{\mathrm{p}}$ with $\Gamma_{\mathrm{p}}\leq 0$, i.e. they have a finite lifetime. According to the result of the previous section, we expect that $\Omega_{\mathrm{p}}\sim r\gamma$ and $\Gamma_{\mathrm{p}}\sim r^2\gamma$, indicating that the quality factor of these excitations scales as $Q = \Omega_{\mathrm{p}}/2\Gamma_{\mathrm{p}}\sim 1/2r$.

\section{Dynamics for $N\geq 8$ atoms}

Interestingly, the dynamics becomes richer and seemingly more chaotic for more than eight atoms, as evident in Fig. S1 for $r=0.001$ and $\beta =1$. The main reason is that for $N\geq 8$ there are more transitions and vacuum Rabi oscillations that become accessible. To understand it, note that if we start at the $j=N/2$ subspace containing the global ground state $\ket{N/2,-N/2}$, the Zeno Hamiltonian $PHP$ can couple us to DF states in the subspaces $j = N/2-2,~N/2-4,~...$ down to $N/2~\mathrm{mod}~2$ (because $j\geq 0$). That is, the Hamiltonian $PHP$ connects DF states with $j$'s differing by multiples of 2.
\begin{figure}
    \centering
    \includegraphics[trim={0.1cm 21.5cm 0 0},clip, scale = 0.75]{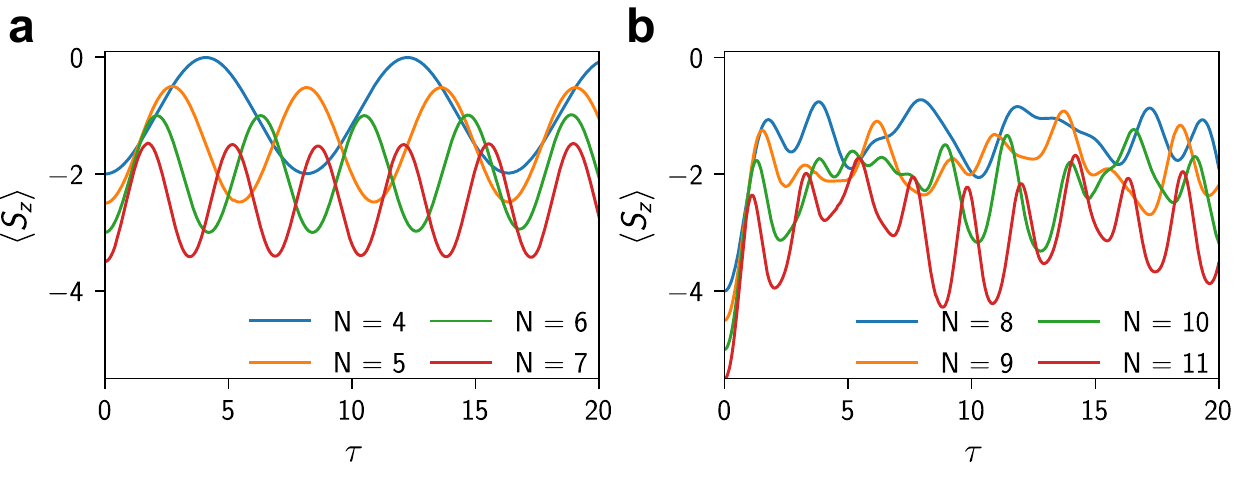}
    \caption{\textbf{Dynamics for different atom numbers.} Atomic population dynamics involving one (\textbf{a}) and multiple (\textbf{b}) Rabi oscillations, for atom number $4\leq N\leq 7$ and $N\geq 8$, respectively.}
   \label{fig:concept}
\end{figure}
For $4\leq N\leq 7$, the Hamiltonian couples the $j=N/2$ ground state to only one other subspace with $j-2 \geq 0$, i.e., two $j$-subspaces in total. For $N\geq8$, the Hamiltonian couples more than two $j$-subspaces: e.g., for $N=8$ we have three such subspaces $j=4,~j-2 = 2,$ and $ j-4 =0$. Furthermore, in each of these subspaces there can be a large DF subspace degeneracy \cite{Paulisch2016UniversalSubspaces}, which may lead to multiple Rabi oscillations and transitions between different subspaces.

\bibliography{references}
\bibliographystyle{ieeetr}